\documentclass{article}
\usepackage{amsmath} 
\usepackage{spconf,graphicx}
\usepackage{booktabs}
\usepackage{ragged2e}
\usepackage{float}
\usepackage[numbers, sort]{natbib} 
\usepackage[
            pdfstartview=FitH,
            CJKbookmarks=true,
            bookmarksnumbered=true,
            bookmarksopen=true,
            linkcolor=red,
            anchorcolor=green,
            citecolor=green
            ]{hyperref}
\usepackage{enumitem}

\setlist{nosep, leftmargin=14pt}

\usepackage{mwe} % to get dummy images

\title{PE-MVCNet:  Multi-view and Cross-modal Fusion Network 
\\for Pulmonary Embolism Prediction}
\name{Author(s) Name(s)\thanks{Some author footnote.}}
\address{Author Affiliation(s)}
\name{
\parbox{\linewidth}{\centering
Zhaoxin Guo$^1$, Zhipeng Wang$^1$, Ruiquan Ge$^{1,7,\star}$, Jianxun Yu$^2$, Feiwei Qin$^{1,\star}$ \\
Yuan Tian$^{3}$, Yuqing Peng$^{4}$, Yonghong Li$^{5}$, Changmiao Wang$^{6}$}
}

\address{$^1$Hangzhou Dianzi University, Hangzhou, China $\,$
$^2$ Dalian Polytechnic University, Dalian, China \\
$^3$ The Second Affiliated Hospital of The Chinese University of Hong Kong, Shenzhen, China\\
$^4$ Shenzhen Institutes of Advanced Technology, Chinese Academy of Sciences, Shenzhen, China\\
$^5$ Shenzhen Institute of Information Technology, China $\,$
$^6$Shenzhen Research Institute of Big Data, China \\
$^7$Hangzhou Institute of Advanced Technology, Hangzhou, China
}
\begin{document}
%\ninept
%
\maketitle
\begin{abstract}

The early detection of a pulmonary embolism (PE) is critical for enhancing patient survival rates. Both image-based and non-image-based features are of utmost importance in medical classification tasks. In a clinical setting, physicians tend to rely on the contextual information provided by Electronic Medical Records (EMR) to interpret medical imaging. However, very few models effectively integrate clinical information with imaging data. To address this shortcoming, we suggest a multimodal fusion methodology, termed PE-MVCNet, which capitalizes on Computed Tomography Pulmonary Angiography imaging and EMR data. This method comprises the Image-only module with an integrated multi-view block, the EMR-only module, and the Cross-modal Attention Fusion (CMAF) module. These modules cooperate to extract comprehensive features that subsequently generate predictions for PE. We conducted experiments using the publicly accessible Stanford University Medical Center dataset, achieving an AUROC of 94.1\%, an accuracy rate of 90.2\%, and an F1 score of 90.6\%. Our proposed model outperforms existing methodologies, corroborating that our multimodal fusion model excels compared to models that use a single data modality. Our source code is available at \href{https://github.com/LeavingStarW/PE-MVCNET}{https://github.com/LeavingStarW/PE-MVCNET}.

\renewcommand{\thefootnote}{\fnsymbol{footnote}}
\footnotetext[1]{Correspondings : gespring@hdu.edu.cn, qinfeiwei@hdu.edu.cn.}
\renewcommand{\thefootnote}{\arabic{footnote}}

\end{abstract}
\begin{keywords}
Multi-view, Cross-modal, Transformer mechanism, CT and EMR data, PE prediction
\end{keywords}
\section{Introduction}
\label{sec:intro}

Pulmonary Embolism (PE), a severe medical condition, is characterized by the blockage of a pulmonary artery due to a blood vessel embolus. This blockage escalates pulmonary vascular resistance and elevates pulmonary artery pressure, placing PE second only to myocardial infarction and sudden death in terms of severity. Timely diagnosis and treatment can reduce the patient's mortality rate to approximately 10\%. Computed Tomography Pulmonary Angiography (CTPA) is predominantly employed as the primary diagnostic technique for PE, as it offers detailed visualization of the thrombus morphology within the patient's pulmonary arteries. However, CTPA images, often numbering in the hundreds for each patient, are vulnerable to variations in imaging technology. These variations present significant challenges for physicians during interpretation, potentially leading to missed diagnoses.

Recent research has extensively applied deep convolutional neural networks \cite{khachnaoui2022deep,grenier2023deep,chen2024scunet++} and attention mechanisms \cite{suman2021attention} to enhance the accuracy of PE diagnosis. Concurrently, techniques such as CNN-LSTM \cite{huhtanen2022automated,shi2020automatic,suman2021attention} have been utilized to consider the relationships between consecutive Computed Tomography (CT) slices, thereby better capturing dependencies among these slices. The most sophisticated model to date, PENet \cite{huang2020penet}, is an end-to-end 3D CNN that leverages multiple CT slices for PE detection. The use of 3D convolutions allows the network to incorporate information from multiple slices during prediction, making the network's ability to learn global information crucial. This is because the presence of PE is not confined to a single CT slice.

Despite the proliferation of deep learning-based methods in the field of medical imaging, a significant issue persists, namely the neglect of how clinicians frequently employ multimodal data for collaborative decision-making in diagnosing clinical conditions. This is due to the fact that data from different modalities can enhance each other. In response to this, Tang et al.\cite{tang2022matr} proposed an unsupervised method that employs a Multiscale Adaptive Transformer to integrate medical image models from two modalities. This method has shown superior performance and generalization ability. Furthermore, the integration of Electronic Medical Record (EMR) data with Computed Tomography (CT) images may present a promising approach. Zhou et al.\cite{zhou2021radfusion} introduced a multimodal fusion model that combines CT and EMR data for the automated classification of Pulmonary Embolism (PE) cases. Comprised of a CT imaging model, an EMR model, and a multimodal fusion model, their work evidenced the superiority of the multimodal model over-reliance on a single data modality.

However, existing methods grapple with issues such as unidimensional data and incomplete multimodal fusion features. To surmount these challenges, our study presents a novel multimodal PE detection framework based on multi-view and cross-modal techniques. Specifically, we deployed a multi-view approach for three-dimensional image feature extraction and prediction. Furthermore, a MLP network with a transformer encoder was implemented to extract and predict features from EMR data. The features extracted from both components were integrated into a cross-modal module, enabling comprehensive feature fusion for the ultimate PE prediction output. The contributions of our method can be encapsulated as follows:

\begin{enumerate}
\item We leverage spatial and dimensional attention to extract pertinent information from CT images from spatial, channel, and dimensional perspectives. 
\item We employ a cross-modal module to learn and align complementary information between two modalities, thereby enhancing the accuracy and robustness of our model by integrating image and tabular features.
\end{enumerate}

\begin{figure*}[t]
    \centering
    \includegraphics[width=0.75\linewidth]{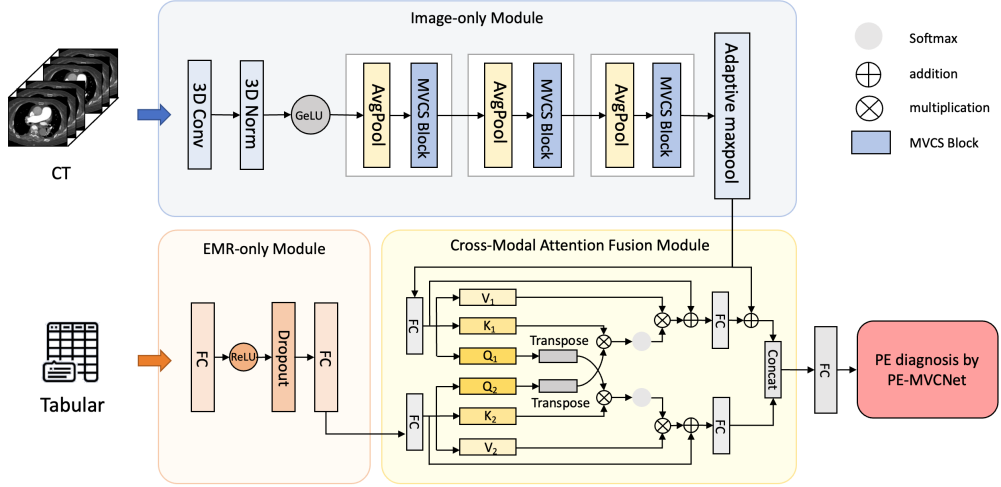}
    \caption{The overall framework of the proposed PE-MVCNet model for PE prediction. The model comprises the Image-only module, EMR-only module, and Cross-modal Attention Fusion (CMAF) module. The Image-only model employs spatial and dimensional attention to investigate dependency relationships on spatial, channel, and dimensional aspects, respectively. Conversely, the CMAF module is designed to capture the correlation between image and tabular features.}
\label{fig:model}
\end{figure*}

\section{Method}
\label{sec:format}
\subsection{Overview of the architecture}
Our study is designed to forecast the presence or absence of PE in a patient through the integration of the patient's chest CTPA image and the corresponding EMR attribute information. This objective is consequently translated into a binary classification task. In this section, we delineate three key elements of our framework: the image-only model, the EMR-only model, and the multimodal fusion module. The architecture of the model is depicted in Figure \ref{fig:model}.

\subsection{Image-only process module}
\begin{figure}[h]
    \centering
    \includegraphics[width=0.6\linewidth]{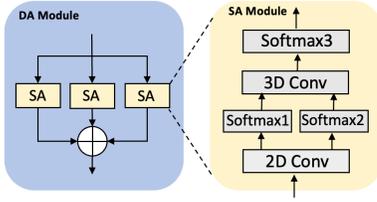}
    \caption{ Multi-View Coupled Self-Attention Block. 'DA' denotes Dimensional Attention, and 'SA' signifies Spatial Attention.}
\label{fig:roc}
\end{figure}

In the domain of medical image classification, the amalgamation of both global and local features within an image is instrumental to the successful categorization of 3D medical images. Conventional network architectures often have a restricted receptive field of the convolutional kernel. Consequently, we crafted our image-only model based on the Multi-View Coupled Self-Attention (MVCS) module proposed by Zhou et al.\cite{zhu2022multi}, the architecture of which is illustrated in Figure \ref{fig:model}. Specifically, the MVCS Block incorporates spatial and dimensional attention mechanisms into the 3D ResNet \cite{he2016deep}. These dual attention mechanisms can capture both global and local information of the CT image across three dimensions, systematically modeling the correlations among the space, channel, and dimension. Figure \ref{fig:roc} displays the specific structure of the MVCS block.

\textbf{Spatial Attention } The input $X$ is initially transformed into three distinct views: $X^0 \in R^{BD\times H \times W \times C}$, $X^1 \in R^{BH \times W \times D \times C}$, and $X^2 \in R^{BW \times H \times D \times C}$, wherein $B$ denotes the batch size, $C$ signifies the number of channels, while $W$, $H$, and $D$ represent the width, height, and number of slices, respectively. Each view is subsequently mapped to a key, query, and value using a 1x1 convolution. The results are expressed as ${X}_k^t$, $X_q^t$, and $X_v^t$, where $t$ signifies the view index.

For the spatial attention mechanism, each view yields corresponding matrices $X_k^t$ and $X_q^t$, which are subsequently reshaped into $HW \times C^\prime$\ and $C^\prime \times HW$ respectively. The spatial similarity matrices $M_S^t\in R^{HW\times HW}$ are then formulated through $X_q^t \times X_k^t$. This approach effectively captures distant dependencies in the spatial dimension. Similarly, the channel similarity matrix $M_C^t\in R^{C^\prime\times C^\prime}$ is formulated through $X_k^t \times X_q^t$, thereby capturing remote dependencies in the channel dimension.

\textbf{Dimensional Attention } In order to extract the remote relationships between slices more comprehensively, we utilize a dimensional attention mechanism, which is appended after the spatial attention. The input $X$ is mapped into the spatial key, query, and value through a 3$\times$1$\times$1 convolution, denoted as $X_k\in R^{B\times D\times H\times W\times C}$, $X_q\in R^{B\times D\times H\times W\times C}$, and $X_v\in R^{B\times D\times H\times W\times C}$. Post-mapping, $X_q$ and $X_k$ are reshaped into matrices that are suitable for computation. Subsequently, these two matrices are multiplied to generate a similarity matrix $M_D^t\in R^{D\times D}$ along the third dimension. This matrix signifies the degree of correlation among different slices. The final output features of view $t$ can be articulated as follows:

\begin{footnotesize}
\begin{equation}
\mathrm{X} = \sum_{t=0}^2\left(\operatorname{softmax} \left(M_S^t\right)+\text { softmax }\left(M_C^t\right)
              +\text { softmax }\left(M_D^t\right)\right) \times X_v^t,
\end{equation}
\end{footnotesize}
where $M_S^t$, $X_q^t$, and $M_D^t$ denote the spatial, channel, and dimensional similarity matrices, respectively.

\subsection{EMR-only process module}

In order to extract features from the EMR data, we began by normalizing the six form files. This involved removing features with zero variance and adjusting the remaining attributes by subtracting their mean and dividing by their standard deviation. Subsequently, the data was consolidated into a single table, based on the patients' index numbers, which served as the input for the Electronic Medical Records (EMR)-only model. 

We first conduct dimensionality reduction using LinearSVC on the EMR data, then use TabNet \cite{arik2021tabnet} to transform the data into suitable embeddings, which serve as inputs for the MLP. Particularly, TabNet is not involved in the overall training, only used for data transformation. And we utilized a simple MLP network, as illustrated in Figure \ref{fig:model}. This network is composed of fully connected layers, dropout layers, and ReLU activation functions. The hierarchical structure of this network introduces nonlinearity into the model, thereby improving its adaptability to complex EMR data. The choice of this structure was driven by the intent to enhance the model's utilization of attributes, thereby improving the accuracy of predicting the presence of PE in patients. 

\subsection{Cross-modal fusion module}
To explore the inherent correlations between images and EMR data, we incorporated a cross-modal module into our study. To be specific, features were independently extracted from the imaging-only model and the EMR-only model. These were then fed into the CMAF module \cite{luo2022cmafgan} to facilitate comprehensive feature fusion. 

Within the CMAF module, the inputs comprised image features $x_i$ and text features $y_i$. Owing to GPU memory capacity constraints, we initially transformed these two features into $x\in R^{B \times D}$ and $y\in R^{B \times D}$ by employing a fully connected layers, thereby mapping them into two different feature spaces. Subsequently, the degree of match was computed as follows:
\begin{equation}
\begin{gathered}
    \beta_{j, i}=\frac{\exp \left(s_{i j}\right)}{\sum_{i=1}^S \exp \left(s_{i j}\right)}, \text { where } s_{ij}=q_1\left(x_i\right)^T k_2\left(y_j\right), \\
    \rho_{j, i}=\frac{\exp \left(t_{i j}\right)}{\sum_{i=1}^S \exp \left(t_{i j}\right)}, \text { where }  t_{ij}=q_2\left(y_i\right)^T k_1\left(x_j\right),
\end{gathered}
\end{equation}
where $S = W \times H$, and $\beta$ and $\rho$ represent the matching degree within the image and text spaces, respectively. Subsequently, the calculated $\beta_{j, i}$ and $\rho_{j, i}$ are multiplied with the feature values to generate the final cross-modal attention maps. Finally, these maps are combined with the image features $x_i$ and fed into subsequent fully connected layers, which ultimately generate predictions for PE.

\section{Experiment}
\label{sec:pagestyle}
\subsection{Experimental setting}

\textbf{Dataset.} We utilized a publicly accessible dataset provided by Stanford University \cite{zhou2021radfusion}. This dataset comprises 1,837 axial CTPA exams from 1,794 patients, spanning from 2000 to 2016, with a CT slice thickness of 1.25mm for each patient. Corresponding EMR data is also available for each patient. The dataset exhibits a near-equal distribution of positive and negative PE labels. These labels are presented in a list form, with '0' denoting negative PE and '1' indicating positive PE. All labels were generated via a manual review conducted by a board-certified radiologist. The EMR data encompasses multiple tables, including demographics, vital signs, inpatient medications, outpatient medications, ICD codes, and laboratory test results. We processed the EMR data as delineated in Section 2.3 and consolidated these structured EMRs into a single tabular file, which complements the dataset. To ensure a fair comparison, we adhered to the standard split from PEFusion \cite{zhou2021radfusion}, with the training, validation, and testing splits set at 80\%, 10\%, and 10\%, respectively. We guaranteed that no patient overlap occurred between each subset. 

% \begin{table}[h]
% \caption{Data characteristics of the dataset.}
% \label{dataset}

% \centering
% \scalebox{0.85}{
%     \begin{tabular}{ccccc}
%     \toprule
%      & Overall & Train & Validation & Test \\
%     \midrule
%     Number of studies & 1,797 & 1,461 & 167 & 169   \\
%     Number of positive PE & 655 & 488 & 82 & 85 \\
%     Number of negative PE & 1,142 & 973 & 85 & 84 \\
%     \bottomrule
%     \end{tabular}
% }
% \end{table}

\justifying 
\noindent \textbf{Implementation Details. } Experiments were conducted using two NVIDIA HGX A100 Tensor Core GPUs. The SGD optimizer was deployed for this process. The training epoch, learning rate, and batch size were set at 200, 0.01, and 128, respectively. 

\subsection{Comparison with state-of-the-art models}
\justifying
Our proposed model was compared with state-of-the-art methods, which included single-modality strategies such as 3D ResNet50 \cite{he2016deep}, 3D ResNet101 \cite{he2016deep}, PENet \cite{huang2020penet}, and a multimodal fusion model, PEfusion \cite{zhou2021radfusion}. The same dataset was utilized for all models. The results of our proposed method and the comparative methods are depicted in Table \ref{comparison}. 

\begin{table}[h]
\caption{Comparison with state-of-the-art models.}
\label{comparison}
%\centering
\scalebox{0.6}{
    \begin{tabular}{cccccccc}
    \toprule
    Methods & AUROC & ACC & F1 score & Specificity & Sensitivity & PPV & NPV \\
    \midrule
    3D ResNet50 \cite{he2016deep} & 0.694 & 0.556 & 0.687 & 0.785 & 0.963 & 0.534 & 0.785  \\
    3D ResNet101 \cite{he2016deep} & 0.722 & 0.611 & 0.701 & 0.757 &  0.902 & 0.574 & 0.757 \\
    PENet \cite{huang2020penet} & 0.660 & 0.623 & 0.666 & 0.656 & 0.743 & 0.604 & 0.656\\
    PEfusion \cite{zhou2021radfusion} & 0.936 & 0.882 & 0.882 & 0.900 & 0.866 & 0.898 & 0.867 \\
    \textbf{PE-MVCNet(Ours)} & \textbf{0.941} & \textbf{0.902} & \textbf{0.906} & \textbf{0.932} & \textbf{0.939} & \textbf{0.899} & \textbf{0.932} \\
    \bottomrule
    \end{tabular}
}
\end{table}

\justifying
From Table \ref{comparison}, our model emerges superior across all metrics when compared to other state-of-the-art methods. Specifically, in comparison with the single-modality method, our method enhances the Area Under the Receiver Operating Characteristic(AUROC) by up to 0.281, increases the accuracy by 0.346, and boosts the F1 score by 0.240. These improvements suggest that our multimodal approach effectively amalgamates the interrelations between image and text data, compared to models that rely solely on a single data modality. Utilizing two modalities as inputs not only offers a comprehensive interpretation of the data but also optimizes the complementarity between different modalities. When compared to PEfusion, our model exhibits an increase in AUROC, accuracy, and F1 score by 0.005, 0.020, and 0.024, respectively. This underscores our model's proficiency in feature fusion. The introduced CMAF module adeptly captures the inherent correlations between the two modalities, thereby providing the model with richer information. 

% \begin{figure}[h]
%     \centering
%     \includegraphics[width=0.7\linewidth]{images/roc1.png}
%     \caption{ ROC curves between ours and other methods.}
% \label{fig:roc}
% \end{figure}

\subsection{Ablation study}
To validate the effectiveness of the multi-view module and the cross-modal module, we carried out ablation experiments. These experiments involved an image-only model, an EMR-only model, and a model without the CMAF module. The results of these experiments are presented in Table \ref{ablation}.

\begin{table}[h]
\caption{ Ablation studies.}
\label{ablation}
%\centering
\scalebox{0.62}{
    \begin{tabular}{cccccccc}
    \toprule
    Methods & AUROC & ACC & F1 score & Specificity & Sensitivity & PPV & NPV \\
    \midrule
    Image-only & 0.699 & 0.630 & 0.590 & 0.602 & 0.524 & 0.672 & 0.602  \\
    EMR-only & 0.902 & 0.890 & 0.905 & 0.873 & 0.909 & 0.901 & 0.873 \\
    Without CMAF & 0.936 & 0.895 & 0.900 & 0.931 & 0.939 & 0.865 & 0.932 \\
    \textbf{PE-MVCNet(Ours)} & \textbf{0.941} & \textbf{0.902} & \textbf{0.906} & \textbf{0.932} & \textbf{0.939} & \textbf{0.899} & \textbf{0.932} \\
    \bottomrule
    \end{tabular}
}
\end{table}

Table \ref{ablation} distinctly demonstrates that our fusion model significantly outperforms the two most effective single-modality models. In particular, the fusion model exhibits an increased AUROC by 0.242 and 0.039 compared to the image-only and EMR-only models, respectively. In addition, the model's accuracy is superior by 0.272 and 0.012, respectively, while its F1 score is greater by 0.316 and 0.001, respectively.

In addition, our model exceeds the AUROC, accuracy, and F1 score of the simple fusion model, which does not utilize the CMFA module, by 0.005, 0.007, and 0.006, respectively. This suggests that our cross-modal module can effectively amalgamate feature information from two different modalities. This fusion capability enables the model to gain a comprehensive understanding and utilization of information from different data sources such as images and texts. Consequently, it demonstrates superior performance in predictions.

\section{Conclusion}
\label{sec:majhead}
\justifying
This study aimed to establish a multimodal deep learning model for diagnosing pulmonary embolism by harnessing information from CT images and EMR data. The experimental results demonstrate that our proposed multimodal model excelled with an AUROC of 94.1\%, accuracy of 90.2\%, and an F1 score of 90.6\%, outperforming all other models compared. The improvement in AUROC compared to the image-based model was 24.2\%, the EMR-based model was 3.9\%, and the model lacking the cross-modal module was 0.5\%. Specifically, we elaborated on a multimodal fusion strategy based on multi-view and cross-modal approaches. The multi-view module was designed to extract features from the spatial, channel, and dimensional aspects of CT images, while the cross-modal module effectively integrated features from both CT images and EMR data. The preliminary results indicated considerable improvements in augmenting model performance and robustness compared to single-modal methods. Implementing our approach allowed the model to thoroughly comprehend and utilize information from diverse data sources in a comprehensive manner, thereby providing robust support to enhance the accuracy and reliability of pulmonary embolism detection.

\section{Compliance with Ethical Standards}
This research study was conducted retrospectively using human subject data made available in open access by Stanford University Medical Center (SUMC) \cite{zhou2021radfusion} dataset. Ethical approval was not required as confirmed by the license attached with the open-access data.

\section{Acknowledgements}
This work was supported by the Natural Science Foundation of Zhejiang Province (Nos. LY21F020017, LY21F020015), GuangDong Basic and Applied Basic Research Foundation (No.2022A1515110570), Innovation Teams of Youth Innovation in Science and Technology of High Education Institutions of Shandong Province (No. 2021KJ088), Shenzhen Science and Technology Program (No.KCXFZ20201221173008022). All authors declare that they have no conflicts of interest.

%\pagebreak

\bibliographystyle{IEEEbib}
% \bibliography{refs}

% Generated by IEEEtran.bst, version: 1.14 (2015/08/26)

\end{document}